# A novel approach to the synthesis of the electromagnetic field distribution in a chain of coupled resonators

M.I. Ayzatsky[1]

National Science Center Kharkov Institute of Physics and Technology (NSC KIPT),
610108, Kharkov, Ukraine

A novel approach to the synthesis of the electromagnetic field distribution in a chain of coupled resonators has been developed. This approach is based on the new matrix form of the solutions of the second-order difference equations. If a chain of coupled resonators can be described by the second-order difference equation for amplitudes of expansion of the electromagnetic field, two linearly independent solutions can be constructed on the basis of the solutions of nonlinear Riccaty equation. Setting the structure of one solution, from the Riccaty equation we can find the electrodynamical characteristics of resonators and coupling holes, at which the desired distribution of amplitudes is realized. On the base of this approach we considered the problem of separation of the electromagnetic field into "forward" and "backward" components in the inhomogeneous chain of resonators. It was shown that in the frame of considered model such separation is not defined uniquely.

## 1 Introduction

There are three main fields of using the coupled resonator chains – accelerators [1], RF-sources, mainly travelling wave tubes (TWT) [2] and RF filters [3]. If for the first two applications it is necessary to create the special field distribution for the given frequency (accelerators) or some frequency range (TWT) along of the chain, then for the RF filters requirements are imposed on the amplitude-frequency and phase-frequency characteristics at the chain output.

Coupled-resonator circuits are of importance for design of RF/microwave filters, in particular, the narrow-band bandpass filters that play a significant role in many applications. There is a general technique for designing coupled-resonator filters in the sense that it can be applied to any type of resonator despite its physical structure. [4,5]

In coupled-cavity TWTs several tens of coupled cavities are used as the slow wave structure. The efficiency of a TWT is limited by peculiarity of the bunching process and the bunch transfer from decelerating phase into the accelerating phase of the RF field. The usual technique suggested for increasing the efficiency involves tapering of the wave phase velocity so that the decelerated bunches remain within the decelerating phase of the wave. There were proposed several methods for synthesis of the optimum phase velocity distribution along the slow wave structure (see, for example, [6,7,8,9,10,11,12,13]).

The widest use the the cavity chains have found in the accelerator technique. At the very beginning of its development, the RF accelerators have the RF resonators as the main element of its construction. Disk-loaded waveguides [14,15,16,17,18,19], different side-coupling standing wave structures[20,21], hybrid (combined) accelerating structures (the initial part of the structure is a standing wave buncher, and its main part is a disk-loaded waveguide) [22,23] - this is a short enumeration of the different coupled resonator chains that are used in accelerators. There is enormous number of publications that describe the calculation and design the accelerating structures.

---

[1] M.I. Aizatskyi, N.I.Aizatsky; aizatsky@kipt.kharkov.ua

The calculation of parameters and the design play an important role in the process of developing an accelerating structure. No less important role is played by the process of tuning cells after section brazing.

In order to provide synchronism with the beam and electromagnetic field in the accelerating structure, the phase advance of each cell needs to be adjusted to its nominal value. This can be done after brazing by correcting machining deviations, assembly and brazing mismatching. This adjustment process is called tuning (post-tuning).

The tuning methods based on the non-resonant perturbation field distribution measurement [24,25,26,27,28,29,30] have been widely used for tuning travelling-wave structures, especially in tuning the constant-gradient ones. [31,32,33,34,35,36,37,38,39,40,41,42,43,44].

There are several approaches for post-tuning. The most widespread tuning method became one, in which the field distribution was considered to be a linear superposition of forward and backward waves in each cell [31]. The internal reflection of each cell was obtained by calculating the difference of the amplitudes of the backward waves seen before and after that cell. But forward and backward waves were not strictly determined. Their amplitudes were introduced phenomenologically.

Development of the Coupling Cavity Model (CCM) [45,46,47,48]] gives possibility to look into this method more deeply [49,50]. However, the problem of expanding the electromagnetic field into the forward and backward waves in each cell of the inhomogeneous chain has not been cleared up yet.

In this article a novel approach to analysis of the electromagnetic field distribution in a chain of coupled resonators is presented. This approach is based on the new matrix form of the solutions of the second-order difference equations [51].

## 2 Second-order linear difference equation for the chain of the finite number of resonators

In the frame of the Coupling Cavity Model (CCM) electromagnetic field in each cavity of the chain of resonators are represented as the expansion with the short-circuit resonant cavity modes [17,18,52,53,54,55]

$$\vec{E}^{(k)} = \sum_q e_q^{(k)} \vec{E}_q^{(k)}(\vec{r}) ,  \qquad (1)$$

where $q = \{0, m, n\}$ and such coupling equations for $e_{010}^{(n)}$ can be obtained [45,47,48]

$$Z_k e_{010}^{(k)} = \sum_{j=-\infty, j \neq n}^{\infty} e_{010}^{(j)} \alpha_{010}^{(k,j)} . \qquad (2)$$

Here $e_{010}^{(k)}$ - amplitudes of $E_{010}$ modes, $Z_k = 1 - \dfrac{\omega^2}{\omega_{010}^{(k)2}} - \alpha_{010}^{(k,k)}$, $\omega_{010}^{(k)}$ - eigen frequencies of these modes, $\alpha_{010}^{(k,j)}$ - real coefficients that depend on both the frequency $\omega$ and geometrical sizes of all volumes. Sums in the right side can be truncated

$$Z_k^{(N)} e_{010}^{(N,k)} = \sum_{j=k-N, j \neq k}^{k+N} e_{010}^{(N,j)} \alpha_{010}^{(k,j)} . \qquad (3)$$

In the case of $N = 1$, the system of coupled equations (3) is very similar to the one that can be constructed on the basis of equivalent circuits approach (see, for example [20,56,57,58]). But in the frame of the CCM the coefficients $\alpha_{0mn}^{(k,j)}$ are electrodynamically strictly defined for arbitrary $N$ and can be calculated with necessary accuracy. In the theory of RF filters the coupling matrix circuit model is used intensively (see, for example, [59] and cited there literature). The main problem is how to calculate the matrix elements.

Amplitudes of other modes ($(m,n) \neq (1,0)$) can be found by summing the relevant series

$$e_{0mn}^{(k)} = \frac{\omega_{0mn}^{(k)2}}{\omega_{0mn}^{(k)2} - \omega^2} \sum_{j=k-N}^{k+N} e_{010}^{(j)} \alpha_{0mn}^{(k,j)}. \tag{4}$$

For the chain of cylindrical resonators longitudinal component of electric field at $r = 0$ (on the system longitudinal axis) is:

$$E_z^{(k)} = \sum_{m,n} e_{0mn}^{(j)} \cos\left(\frac{\pi}{d} n z\right). \tag{5}$$

If we can ignore "long coupling" interaction, the set of coupling equations (3) takes the form[2]

$$Z_k e_{010}^{(k)} = e_{010}^{(k-1)} \alpha_{010}^{(k,k-1)} + e_{010}^{(k+1)} \alpha_{010}^{(k,k+1)}, \tag{6}$$

where $Z_k = \left(1 - \frac{\omega^2}{\omega_{010}^{(k)2}} - \alpha_{010}^{(k,k)} - i\frac{\omega}{\omega_{010}^{(k)} Q_k}\right)$

The set of coupling equations (6) can be considered as the second-order difference equation. This difference equation, which defines the amplitudes of the basic modes $e_{010}^{(k)}$, is the main equation of the CCM. It is reasonable to note that the amplitudes of the basic modes $e_{010}^{(k)}$ are non-measured values. Indeed, we can measure the components of electric field in any point, for example, by the nonresonant perturbation method, but we cannot measure $e_{0mn}^{(k)}$ and have to use numerical methods for finding these amplitudes by using the expansion (1). This circumference create difficulties in studding the properties of the real slow-wave waveguides, including their tuning [31,49]. The similar situation arises also in other electrodynamic models. For example, the space harmonics in homogeneous periodic waveguides are non-measured values, too.

We will consider the chain with the finite number of resonators (see Figure 1). The first and the last resonators are connected to the transmission lines[3] and the equations (6) for $k = 1$ and $k = N$ have to be changed [58]

$$Z_1 e_{010}^{(1)} = \left[-\frac{\omega^2}{\omega_{010}^{(1)2}} - i\frac{\omega(1+\beta_1)}{\omega_{010}^{(1)} Q_1} + (1 - \alpha_{010}^{(1,1)})\right] e_{010}^{(1)} = \alpha_{010}^{(1,2)} e_{010}^{(1)} + \frac{2i\omega}{\omega_{010}^{(1)} Q_1} \sqrt{\frac{\beta_1 R_1}{Z}} \frac{1}{d_1} U \tag{7}$$

$$Z_N e_{010}^{(N)} = \left[-\frac{\omega^2}{\omega_{010}^{(N)2}} - i\frac{\omega(1+\beta_N)}{\omega_{010}^{(N)} Q_N} + (1 - \alpha_{010}^{(N,N)})\right] e_{010}^{(N)} = \alpha_{010}^{(N,N-1)} e_{010}^{(N-1)} \tag{8}$$

where

$\beta_1$, $\beta_N$ - coupling factors of the first and the last resonators with transmission lines,

$Z$ - impedance of the input transmission line

$R_1$ - shunt impedance of the first resonator

$U = \sqrt{PZ/2}$, $P$ - power of the external RF source

$d_k$ - length of the $k$-th resonator

Amplitude of the reflected wave in the input transmission line is

$$U_R = -\sqrt{\frac{\beta_1 Z}{R_1}} d_1 e_{010}^{(1)} - U \tag{9}$$

---

[2] There is a problem of taking into account absorption of RF energy in walls as there are difficulties in obtaining appropriate eigen functions for cylindrical regions. All developed procedures in the frame of the CCM do not include this phenomenon. We used the simplest approach for including absorption into consideration. We supposed that the coupling coefficient do not depend on absorption and include the quality factor into the resonant term in the equations for $e_{010}$ amplitudes.

[3] We will consider the chains with the transmission lines connected to the first and end resonators. Other connections can be considered similarly.

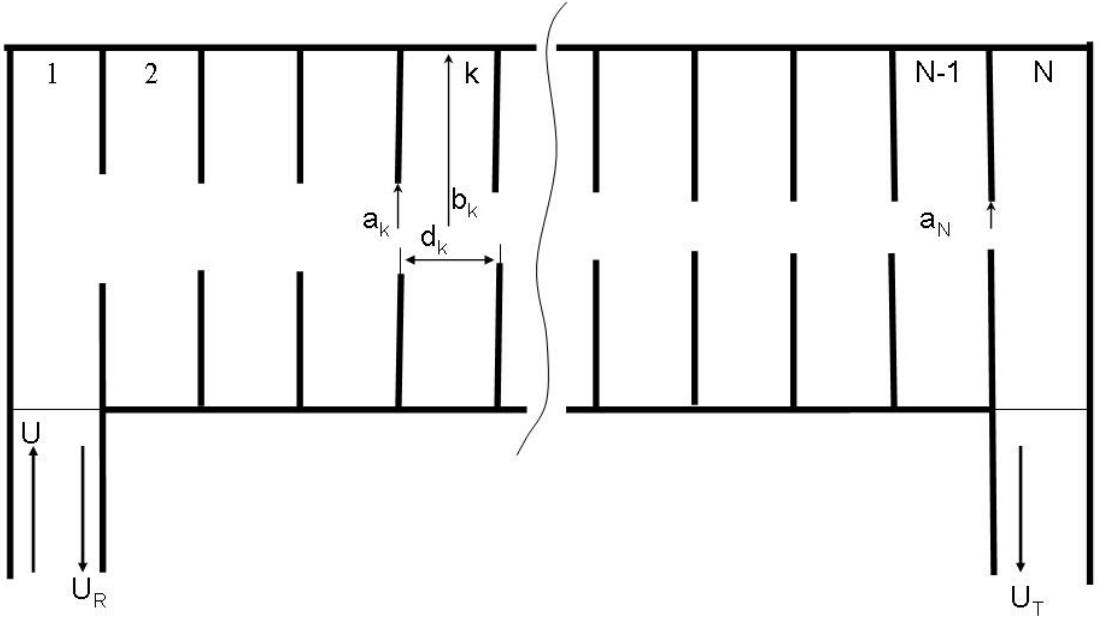

Figure 1

According to the results of the work [60], we will seek a solution of difference equations (6), (7), (8) as

$$e_{010}^{(k)} = y_k^{(1)} + y_k^{(2)}, 1 \leq k \leq N$$
$$e_{010}^{(k+1)} = \rho_k^{(1)} y_k^{(1)} + \rho_k^{(2)} y_k^{(2)}, 1 \leq k \leq N-1, \; \rho_k^{(1)} \neq \rho_k^{(2)} \quad (10)$$

Using this representation, the equations (7),(8) can be rewritten as

$$\left(Z_1 - \alpha_{010}^{(1,2)} \rho_1^{(1)}\right) y_1^{(1)} + \left(Z_1 - \alpha_{010}^{(1,2)} \rho_1^{(2)}\right) y_1^{(2)} = \frac{2i\omega}{Q_1 \omega_{010}^{(1)}} \sqrt{\frac{\beta_1 R_1}{Z}} \frac{1}{d_1} U \quad (11)$$

$$\left(Z_N \rho_{N-1}^{(1)} - \alpha_{010}^{(N,N-1)}\right) y_{N-1}^{(1)} + \left(Z_N \rho_{N-1}^{(2)} - \alpha_{010}^{(N,N-1)}\right) y_{N-1}^{(2)} = 0 \quad (12)$$

and there are such matrix difference equation for new unknowns [51]

$$\begin{pmatrix} y_{k+1}^{(1)} \\ y_{k+1}^{(2)} \end{pmatrix} = T_k \begin{pmatrix} y_k^{(1)} \\ y_k^{(2)} \end{pmatrix}, 1 \leq k \leq N-2. \quad (13)$$

where

$$T_k = \begin{pmatrix} -\dfrac{\left\{\alpha_{010}^{(k+1,k)} - \left(Z_{k+1} - \alpha_{010}^{(k+1,k+2)} \rho_{k+1}^{(2)}\right) \rho_k^{(1)}\right\}}{\alpha_{010}^{(k+1,k+2)} \left(\rho_{k+1}^{(1)} - \rho_{k+1}^{(2)}\right)} & -\dfrac{\left\{\alpha_{010}^{(k+1,k)} - \left(Z_{k+1} - \alpha_{010}^{(k+1,k+2)} \rho_{k+1}^{(2)}\right) \rho_k^{(2)}\right\}}{\alpha_{010}^{(k+1,k+2)} \left(\rho_{k+1}^{(1)} - \rho_{k+1}^{(2)}\right)} \\ \dfrac{\left\{\alpha_{010}^{(k+1,k)} - \left(Z_{k+1} - \alpha_{010}^{(k+1,k+2)} \rho_{k+1}^{(1)}\right) \rho_k^{(1)}\right\}}{\alpha_{010}^{(k+1,k+2)} \left(\rho_{k+1}^{(1)} - \rho_{k+1}^{(2)}\right)} & \dfrac{\left\{\alpha_{010}^{(k+1,k)} - \left(Z_{k+1} - \alpha_{010}^{(k+1,k+2)} \rho_{k+1}^{(1)}\right) \rho_k^{(2)}\right\}}{\alpha_{010}^{(k+1,k+2)} \left(\rho_{k+1}^{(1)} - \rho_{k+1}^{(2)}\right)} \end{pmatrix} \quad (14)$$

Values of the grid vectors in the first and the (N-1)-th cells are connected by a linear relation

$$y_{N-1}^{(1)} = T_{11}^\Sigma y_1^{(1)} + T_{12}^\Sigma y_1^{(2)}$$
$$y_{N-1}^{(2)} = T_{21}^\Sigma y_1^{(1)} + T_{22}^\Sigma y_1^{(2)} \quad (15)$$

Using these relations, the equations (7) and (8) can be rewritten in the form

$$\left(Z_1 - \alpha_{010}^{(1,2)} \rho_1^{(1)}\right) y_1^{(1)} + \left(Z_1 - \alpha_{010}^{(1,2)} \rho_1^{(2)}\right) y_1^{(2)} = \frac{2i\omega}{Q_1 \omega_{010}^{(1)}} \sqrt{\frac{\beta_1 R_1}{Z}} \frac{1}{d_1} U \quad (16)$$

$$\left[\left(Z_N \rho_{N-1}^{(1)} - \alpha_{010}^{(N,N-1)}\right)T_{11}^{\Sigma} + \left(Z_N \rho_{N-1}^{(2)} - \alpha_{010}^{(N,N-1)}\right)T_{21}^{\Sigma}\right]y_1^{(1)} +$$
$$+\left[\left(Z_N \rho_{N-1}^{(1)} - \alpha_{010}^{(N,N-1)}\right)T_{12}^{\Sigma} + \left(Z_N \rho_{N-1}^{(2)} - \alpha_{010}^{(N,N-1)}\right)T_{22}^{\Sigma}\right]y_1^{(2)} = 0 \quad (17)$$

We can choose the sequences $\rho_k^{(1)}$ and $\rho_k^{(2)}$ in such way that the matrix $T_k$ will be the diagonal one [51]. From (14) it follows that $T_{k,12} = T_{k,21} = 0$ for $\rho_k^{(1)}$, $\rho_k^{(2)}$ which fulfilled Riccaty type difference equation (the second-order rational difference equation) [61,62] with different initial values of $\rho_1^{(1)}$ and $\rho_1^{(2)}$ ($\rho_1^{(1)} \neq \rho_1^{(2)}$)

$$\alpha_{010}^{(k+1,k)} - \left(Z_{k+1} - \alpha_{010}^{(k+1,k+2)} \rho_{k+1}\right)\rho_k = 0, \ 1 \leq k \leq N-2 \quad (18)$$

Solution of the matrix difference equation (13) with the diagonal matrix $T_k$ is

$$y_k^{(1)} = \prod_{s=1}^{k-1} \rho_s^{(1)} y_1^{(1)}, \ 2 \leq k \leq N$$
$$y_k^{(2)} = \prod_{s=2}^{k-1} \rho_s^{(2)} y_1^{(2)} \quad (19)$$

We will call $\rho_k^{(1)}$ and $\rho_k^{(2)}$ as characteristic multipliers.

In this case, the equation (17) transforms into

$$\left(Z_N \rho_{N-1}^{(1)} - \alpha_{010}^{(N,N-1)}\right)T_{11}^{\Sigma} y_1^{(1)} + \left(Z_N \rho_{N-1}^{(2)} - \alpha_{010}^{(N,N-1)}\right)T_{22}^{\Sigma} y_1^{(2)} = 0 \quad (20)$$

where

$$T_{11}^{\Sigma} = \prod_{s=1}^{N-2} T_{s,11} = \prod_{s=1}^{N-2} \rho_s^{(1)} \quad (21)$$

$$T_{22}^{\Sigma} = \prod_{s=1}^{N-2} T_{s,22} = \prod_{s=1}^{N-2} \rho_s^{(2)} \quad (22)$$

Solving the equations (16) and (20), we obtain

$$y_1^{(1)} = \frac{2i\omega}{gQ_1 \omega_{010}^{(1)}} \sqrt{\frac{\beta_1 R_1}{Z}} \frac{1}{d_1} U \left(Z_N \rho_{N-1}^{(2)} - \alpha_{010}^{(N,N-1)}\right)T_{22}^{\Sigma} \quad (23)$$

$$y_1^{(2)} = -\frac{2i\omega}{gQ_1 \omega_{010}^{(1)}} \sqrt{\frac{\beta_1 R_1}{Z}} \frac{1}{d_1} U \left(Z_N \rho_{N-1}^{(1)} - \alpha_{010}^{(N,N-1)}\right)T_{11}^{\Sigma} \quad (24)$$

where

$$g = \left(Z_1 - \alpha_{010}^{(1,2)} \rho_1^{(1)}\right)\left(Z_N \rho_{N-1}^{(2)} - \alpha_{010}^{(N,N-1)}\right)T_{22}^{\Sigma} - \left(Z_1 - \alpha_{010}^{(1,2)} \rho_1^{(2)}\right)\left(Z_N \rho_{N-1}^{(1)} - \alpha_{010}^{(N,N-1)}\right)T_{11}^{\Sigma} \quad (25)$$

We introduced the two linearly independent grid functions $y_k^{(1)}$, $y_k^{(2)}$ which are the product of multipliers $\rho_k^{(1)}$ and $\rho_k^{(2)}$ (see (19)). These multipliers are the solutions of the nonlinear difference equation (18) with different initial values of $\rho_1^{(1)}$ and $\rho_1^{(2)}$. These initial values of $\rho_1^{(1)}$ and $\rho_1^{(2)}$ can be chosen arbitrarily. Therefore, we have a continuous set of the two linearly independent grid functions $y_k^{(1)}$, $y_k^{(2)}$, sum of which gives the same grid function $e_{010}^{(k)} = y_k^{(1)} + y_k^{(2)}$ for the given structure of the chain. In the process of synthesis we can change the structure of the chain in such way that $\rho_k^{(1)}$, $\rho_k^{(2)}$ and $y_1^{(1)}$, $y_1^{(2)}$ will take the required values and the desired electromagnetic field distribution ($e_{010}^{(k)}$) in a chain of coupled resonators will be realized.

It is a usual requirement to insure no reflected signal in steady-state, which corresponds to the matching the input transmission line to the considered chain

$$U_R = -\sqrt{\frac{\beta_1 Z}{R_1}} d_1 \left(y_1^{(1)} + y_1^{(2)}\right) - U = 0. \qquad (26)$$

Substituting (23) and (24) into (26), we obtain

$$\beta_1 = \frac{Q_1 \omega_{010}^{(1)}}{2i\omega} g \left[\left(Z_N \rho_{N-1}^{(1)} - \alpha_{010}^{(N,N-1)}\right) T_{11}^{\Sigma} - \left(Z_N \rho_{N-1}^{(2)} - \alpha_{010}^{(N,N-1)}\right) T_{22}^{\Sigma}\right]^{-1} =$$
$$= \frac{Q_1 \omega_{010}^{(1)}}{2i\omega}\left[-Z_1 + \alpha_{010}^{(1,2)} G\right] = \frac{Q_1 \omega_{010}^{(1)}}{2i\omega}\left[\frac{\omega^2}{\omega_{010}^{(1)2}} + i\frac{\omega(1+\beta_1)}{\omega_{010}^{(1)}Q_1} - (1-\alpha_{010}^{(1,1)}) + \alpha_{010}^{(1,2)} G\right]. \qquad (27)$$

where

$$G = \frac{\left(\rho_1^{(1)} - \rho_1^{(2)} G_N\right)}{(1 - G_N)} \qquad (28)$$

$$G_N = \frac{\left(Z_N \rho_{N-1}^{(1)} - \alpha_{010}^{(N,N-1)}\right) T_{11}^{\Sigma}}{\left(Z_N \rho_{N-1}^{(2)} - \alpha_{010}^{(N,N-1)}\right) T_{22}^{\Sigma}} \qquad (29)$$

From (27) it follows that the critical value of the coupling factor $\beta_1$ is

$$\beta_1 = Q_1 \alpha_{010}^{(1,2)} \frac{\omega_{010}^{(1)}}{\omega} \operatorname{Im} G + 1, \qquad (30)$$

and an additional condition is to be fulfilled

$$1 + \alpha_{010}^{(1,1)} - \frac{\omega^2}{\omega_{010}^{(1)2}} - \alpha_{010}^{(1,2)} \operatorname{Re} G = 0 \qquad (31)$$

As $\beta_1$ is a real positive value, then parameter $G$ has a minimal value

$$\operatorname{Im} G > -\frac{\omega}{\omega_{010}^{(1)} Q_1 \alpha_{010}^{(1,1)}} \qquad (32)$$

If the chain has a single input (standing wave structure), we can create the desired field distribution by choosing the values of $\rho_k^{(1)}$, $\rho_k^{(2)}$ and finding the geometrical parameters of resonators and coupling openings from the Riccaty difference equation (18). Characteristics of the first resonator are determined by equations (30) and (31).

If the chain has two ports (traveling wave structure), there is additional possibilities for manipulating with field distribution. We can create the field distribution based on the one solution $y_k^{(1)}$ ($y_k^{(2)} = 0$). In this case the value of amplitude $e_{010}^{(k)}$ equals the value of amplitude $e_{010}^{(k-1)}$ multiplied by the factor $\rho_k^{(1)}$ (quasiperiodic structure):

$$e_{010}^{(k+1)} = \rho_k^{(1)} e_{010}^{(k)} = \prod_{s=1}^{k} \rho_s^{(1)} e_{010}^{(1)} \qquad (33)$$

Such electromagnetic field distribution can be realized if the initial value of the second solution equals to zero

$$y_1^{(2)} = 0 \qquad (34)$$

From (24) it follows that such condition must be fulfilled

$$Z_N \rho_{N-1}^{(1)} - \alpha_{010}^{(N,N-1)} = 0 \qquad (35)$$

This equation determines the characteristics of the last resonator and the value of coupling with the output transmission line

$$1 + \alpha_{010}^{(N,N)} - \frac{\omega^2}{\omega_{010}^{(N)2}} - \frac{\alpha_{010}^{(N,N-1)} \operatorname{Re} \rho_{N-1}^{(1)*}}{\left|\rho_{N-1}^{(1)}\right|^2} = 0 \qquad (36)$$

$$\beta_N = -Q_N \alpha_{010}^{(N,N-1)} \frac{\omega_{010}^{(N)}}{\omega} \frac{\operatorname{Im} \rho_{N-1}^{(1)*}}{\left|\rho_{N-1}^{(1)}\right|^2} - 1 \tag{37}$$

$$\operatorname{Im} \rho_{N-1}^{(1)*} < -\frac{\omega \left|\rho_{N-1}^{(1)}\right|^2}{Q_N \omega_{010}^{(N)} \alpha_{010}^{(N,N-1)}} \tag{38}$$

From (28) and (29) it follows that $G = \rho_1^{(1)}$ and the matching condition (27) takes the form

$$\left[Z_1 - \alpha_{010}^{(1,2)} \rho_1^{(1)}\right] = -\frac{2i\omega}{Q_1 \omega_{010}^{(1)}} \beta_1. \tag{39}$$

The equations (30) and (31) are also simplified

$$\beta_1 = Q_1 \alpha_{010}^{(1,2)} \frac{\omega_{010}^{(1)}}{\omega} \operatorname{Im} \rho_1^{(1)} + 1 \tag{40}$$

$$1 + \alpha_{010}^{(1,1)} - \frac{\omega^2}{\omega_{010}^{(1)2}} - \alpha_{010}^{(1,2)} \operatorname{Re} \rho_1^{(1)} = 0 \tag{41}$$

$$\operatorname{Im} \rho_1^{(1)} > -\frac{\omega}{\omega_{010}^{(1)} Q_1 \alpha_{010}^{(1,2)}} \tag{42}$$

The initial value $y_1^{(1)}$ (23) do not depend on the characteristic multipliers $\rho_k^{(2)}$

$$y_1^{(1)} = \frac{2i\omega}{\left(Z_1 - \alpha_{010}^{(1,2)} \rho_1^{(1)}\right) Q_1 \omega_{010}^{(1)}} \sqrt{\frac{\beta_1 R_1}{Z}} \frac{1}{d_1} U = -\sqrt{\frac{R_1}{\beta_1 Z}} \frac{1}{d_1} U \tag{43}$$

It is important to note that from (37) and (40) it follows that the coupler is not a symmetric element. Only at $Q \to \infty$ the coupler do not reflect from two sides.

### 3 Solutions of the difference equations for the homogeneous chain

Characteristic multipliers $\rho_k^{(1)}$ and $\rho_k^{(2)}$ are the solutions of the nonlinear difference equation (18) with the initial values $\rho_1^{(1)} \neq \rho_1^{(2)}$. In the general case, these initial values can be chosen arbitrary. Input transmission line matching requirement imposes some restrictions (see (32),(38),(42)) on these values.

For the homogeneous chain, the equation (18) takes the form

$$\rho_{k+1} \rho_k - \rho_k \frac{Z}{\alpha_{010}} + 1 = 0 \tag{44}$$

This equation has two stationary points

$$\rho_k = \chi_{1,2} = \frac{Z}{2\alpha_{010}} \pm i\sqrt{1 - \left(\frac{Z}{2\alpha_{010}}\right)^2} = \exp(\pm i\varphi \mp \gamma) \tag{45}$$

The first stationary point is unstable ($|\chi_1|^{-2} > 1$), as the other one is attractive ($|\chi_2|^{-2} < 1$).

The solutions of the equation (44) is [62]

$$\rho_k = \frac{(\rho_1 - \chi_2)\chi_1^k - (\rho_1 - \chi_1)\chi_2^k}{(\rho_1 - \chi_2)\chi_1^{k-1} - (\rho_1 - \chi_1)\chi_2^{k-1}}, \tag{46}$$

If $\rho_1 = \chi_1$, then $\rho_k = \chi_1$, if $\rho_1 = \chi_2$, then $\rho_k = \chi_2$

If we choose $\rho_1^{(1)} = \chi_1$, $\rho_1^{(2)} = \chi_2$, the grid function $y_k^{(1)}$ will correspond to a "forward traveling wave" and $y_k^{(2)}$ to a "backward one".

$$y_k^{(1)} = \chi_1^{k-1} y_1^{(1)}, \ 2 \leq k \leq N$$
$$y_k^{(2)} = \chi_2^{k-1} y_1^{(2)}, \ 2 \leq k \leq N \quad (47)$$

If we choose $\rho_1^{(1)} \neq \chi_1$ and $\rho_1^{(2)} \neq \chi_2$, the grid functions $y_k^{(1)}$ and $y_k^{(2)}$ will correspond to some combinations of the "traveling waves".

$$y_k^{(1)} = \frac{(\rho_1^{(1)} - \chi_2)\chi_1^{k-1} - (\rho_1^{(1)} - \chi_1)\chi_2^{k-1}}{\chi_1 - \chi_2} y_1^{(1)}, \ 2 \leq k \leq N$$

$$y_k^{(2)} = \frac{(\rho_1^{(2)} - \chi_2)\chi_1^{k-1} - (\rho_1^{(2)} - \chi_1)\chi_2^{k-1}}{\chi_1 - \chi_2} y_1^{(2)}, \ 2 \leq k \leq N \quad (48)$$

where $\rho_1^{(1)} \neq \rho_1^{(2)}$.

The sum of these grid functions ($y_k^{(1)} + y_k^{(2)}$) is a grid function that do not depend on $\rho_1^{(1)}$ and $\rho_1^{(2)}$. Indeed,

$$y_k^{(1)} + y_k^{(2)} =$$
$$= \frac{\left[(\rho_1^{(1)} y_1^{(1)} + \rho_1^{(2)} y_1^{(2)}) - \chi_2(y_1^{(1)} + y_1^{(2)})\right]\chi_1^{k-1} - \left[(\rho_1^{(1)} y_1^{(1)} + \rho_1^{(2)} y_1^{(2)}) - \chi_1(y_1^{(1)} + y_1^{(2)})\right]\chi_2^{k-1}}{\chi_1 - \chi_2} \quad (49)$$

Using (23), (24),(25),(46), we obtain
$$(\rho_1^{(1)} y_1^{(1)} + \rho_1^{(2)} y_1^{(2)}) - \chi_2(y_1^{(1)} + y_1^{(2)}) =$$
$$= (\rho_1^{(2)} - \rho_1^{(1)}) \frac{2i\omega}{gQ_1 \omega_{010}^{(1)}} \sqrt{\frac{\beta_1 R_1}{Z}} \frac{1}{d_1} U \frac{\chi_2^{N-3}(\chi_2^2 - 1)}{\chi_1 - \chi_2} \left[Z_N \chi_2 - \alpha_{010}^{(N,N-1)}\right] \quad (50)$$

$$(\rho_1^{(1)} y_1^{(1)} + \rho_1^{(2)} y_1^{(2)}) - \chi_1(y_1^{(1)} + y_1^{(2)}) =$$
$$= (\rho_1^{(2)} - \rho_1^{(1)}) \frac{2i\omega}{gQ_1 \omega_{010}^{(1)}} \sqrt{\frac{\beta_1 R_1}{Z}} \frac{1}{d_1} U \frac{\chi_1^{N-3}(1 - \chi_1^2)}{\chi_1 - \chi_2} \left[Z_N \chi_1 - \alpha_{010}^{(N,N-1)}\right] \quad (51)$$

$$g = \frac{(\rho_1^{(2)} - \rho_1^{(1)})}{\chi_1 - \chi_2} \begin{bmatrix} Z_1 Z_N (\chi_1^{N-1} - \chi_2^{N-1}) - (Z_1 \alpha_{010}^{(N,N-1)} + Z_N \alpha_{010}^{(1,2)})(\chi_1^{N-2} - \chi_2^{N-2}) + \\ + \alpha_{010}^{(1,2)} \alpha_{010}^{(N,N-1)} (\chi_1^{N-3} - \chi_2^{N-3}) \end{bmatrix} \quad (52)$$

We see, that coefficients in (49) near $\chi_{1,2}^{k-1}$ do not depend on $\rho_1^{(1)}$ or $\rho_1^{(2)}$.

To illustrate the above, we consider the homogeneous chain with $\gamma = 0$ ($Z = 2\alpha_{010} \cos(\varphi)$). Characteristics of the first and the last resonators are determined from equations (36),(37),(40),(41).

For the initial values $\rho_1^{(1)} = \chi_1$, $\rho_1^{(2)} = \chi_2$ ($\rho_k^{(1)} = \chi_1$, $\rho_k^{(2)} = \chi_2$) $y_1^{(2)} = 0$ (see (34)), therefore $y_k^{(2)} = 0$ and $y_k = y_k^{(1)} = \exp[i(k-1)\varphi] y_1^{(1)}$. Phasors of the grid functions $y_k$ for different phase shifts are presented in Figure 2. These phasors are the same at the arbitrary selection of the initial values of $\rho_1^{(1)}$ and $\rho_1^{(2)}$.

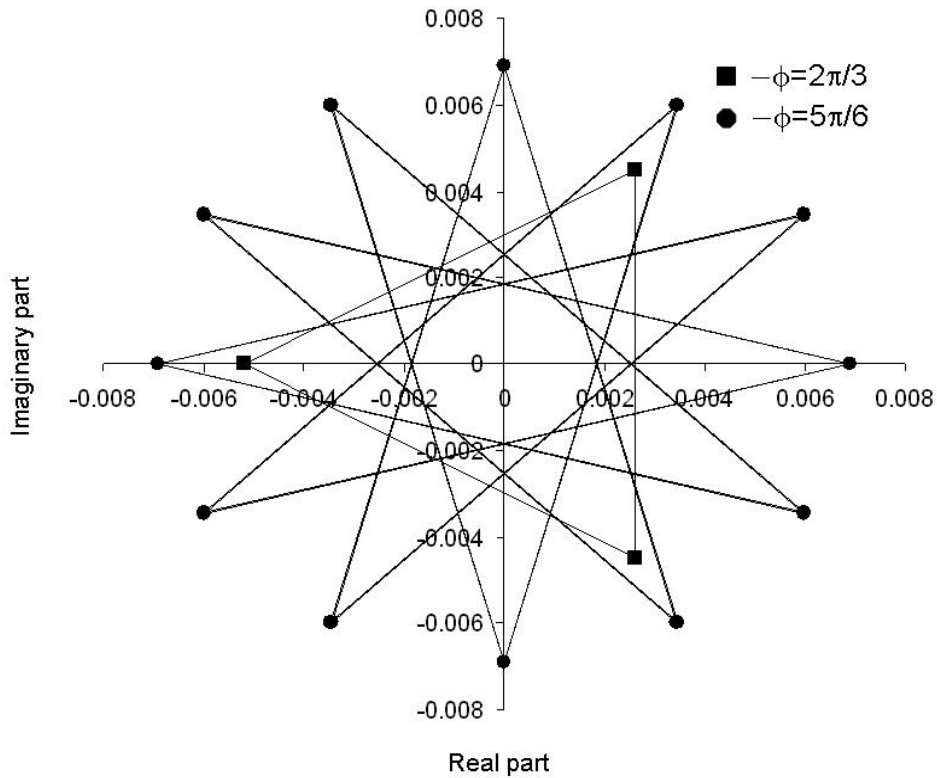

Figure 2 Phasor of the grid function $y_k/q$ for the phase shift per cell $\varphi = 2\pi/3$ and $\varphi = 5\pi/6$; $q = \sqrt{\dfrac{R_1}{Z}}\dfrac{1}{d_*}U$

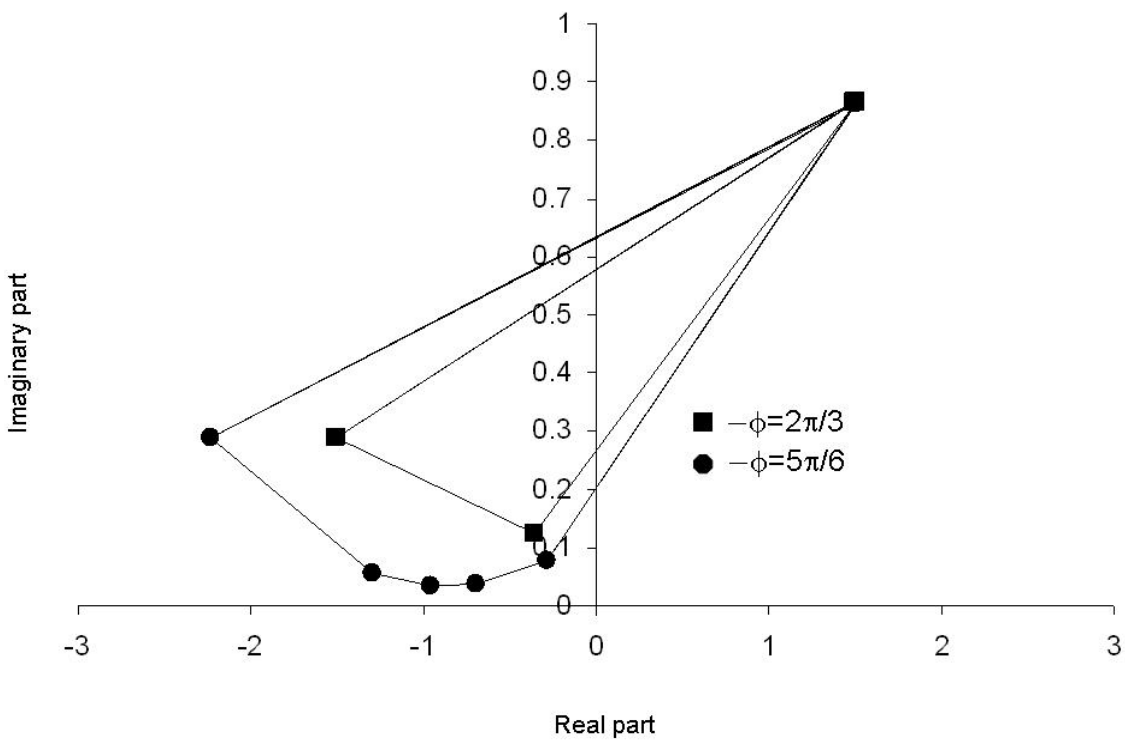

Figure 3 Phasor of the grid function $\rho_k^{(1)}$ for $\rho_1^{(1)} = 2 + \exp(i2\pi/3)$ and for the phase shift per cell $\varphi = 2\pi/3$; $\varphi = 5\pi/6$

As an example, for the initial value $\rho_1^{(1)} = 2 + \exp(i2\pi/3)$ the phasors of the characteristic multipliers $\rho_k^{(1)}$ are presented in Figure 3 and the phasors of the grid function $y_k^{(1)}$ are presented in Figure 4. We can see that the characteristic multipliers $\rho_k^{(1)}$ for selected initial value is subject to a strong variation that leads to the appearance of oscillations in the grid function $y_k^{(1)}$. The same can be said about $\rho_k^{(2)}$ and $y_k^{(2)}$.

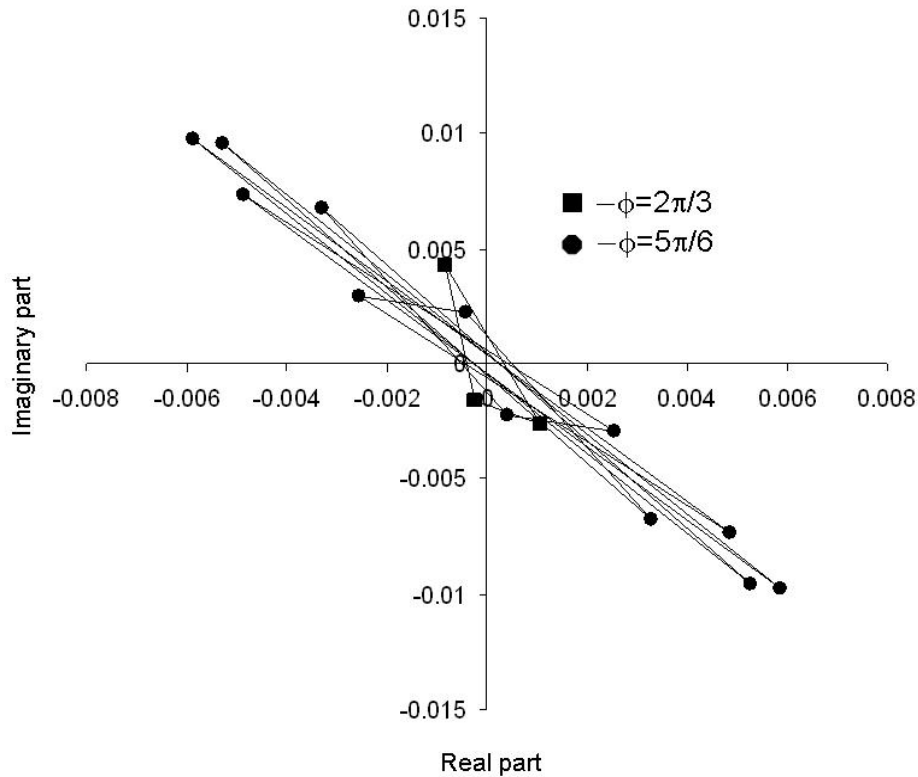

Figure 4 Phasor of the grid function $y_k^{(1)}/q$ for $\rho_1^{(1)} = 2 + \exp(i2\pi/3)$ and the phase shift per cell $\varphi = 2\pi/3$; $\varphi = 5\pi/6$; $q = \sqrt{\dfrac{R_1}{Z}}\dfrac{1}{d_*}U$

## 4 Synthesis of the coupled resonator chain with desired electromagnetic field distribution

In the CCM electromagnetic field distribution is defined by the amplitudes of the basic oscillations [45-48]. For description the lowest passband we have to choose the amplitudes of $E_{010}$ mode as basic oscillations.

In the considered above approach the distribution of amplitude $e_{010}^{(k)}$ is defined by the characteristic multipliers $\rho_k^{(1)}$, $\rho_k^{(2)}$ and initial values of grid functions $y_1^{(1)}$, $y_1^{(2)}$. So, during the synthesis process, we must choose a chain structure (parameters of resonators and coupling elements) such that the coefficients $\rho_k^{(1)}$, $\rho_k^{(2)}$ and $y_1^{(1)}$, $y_1^{(2)}$ will take on the required values and the desired value of amplitudes $e_{010}^{(k)}$ in a chain of coupled resonators will be realized.

Below we will consider the chain of cylindrical resonators that are connected via circular central openings in the walls – the disk loaded waveguides (DLW)[4]. It was shown that the DLWs, that are usually used in linacs, with disk spacing large enough ($d \geq \lambda/3$) can be describe with sufficient accuracy by the difference equation (6) [63]. Appropriate values of the coupling coefficients $\alpha_{010}^{(k,k)}, \alpha_{010}^{(k,k+1)}$ at fixed frequency can be approximated by some functions of geometrical sizes. Calculations on the base of the CCM show that for the most often used in linacs DLWs such approximations can be used

$$\alpha_{010}^{(k,k)} = -\alpha \frac{u_k \overline{p}_k + u_{k+1} \overline{p}_{k+1}}{\tilde{b}_k^2 \tilde{d}_k}$$

$$\alpha_{010}^{(k,k-1)} = \alpha \frac{u_k}{\tilde{b}_k^2 \tilde{d}_k} , \qquad (53)$$

$$\alpha_{010}^{(k,k+1)} = \alpha \frac{u_{k+1}}{\tilde{b}_k^2 \tilde{d}_k}$$

where $u_k = \frac{\alpha a_k^3}{b_*^2 d_*} p_k^{(c)}$, $\tilde{b}_k = \frac{b_k}{b_*}$, $\tilde{d}_k = \frac{d_k}{d_*}$, $a_k$ - the hole radius between $k-1$ and $k$ resonators, $b_k$ - the radius of $k$ cylindrical resonator, $d_k$ - the resonator length, $b_* = c\frac{\lambda_{01}}{\omega}$, $d_*$ - normalizing parameters, $\omega_{010}^{(k)} = c\frac{\lambda_{01}}{b_k}$, $J_0(\lambda_{01}) = 0$, $\alpha = \frac{2}{3\pi J_1^2(\lambda_{01})}$, $\overline{p}_k = \frac{p_k^{(s)}}{p_k^{(c)}}$.

---

[4] DLW structures are the most often used in linacs and represent the chain of cavities in which the phase varies smoothly from cell to cell in such way, that an accelerated particle constantly locates in accelerating field.

Analysis shows that we can consider parameters $p_k^{(s)}, p_k^{(c)}$ as the functions of the geometric sizes of the diaphragms only (the opening radius $a_k$, the thickness $t_k$ of the diaphragm between $k-1$ and $k$ resonators and the radius of the rounding of the disk hole edges).

For $t_k = 0.4$ cm, $d_k = 3.0989$ cm, parameters $p_k^{(s)}, p_k^{(c)}$ can be represented[5] as

$$p_k^{(s)} = 0.0142 a_k^2 - 0.1329 a_k + 0.9133$$
$$p_k^{(c)} = -0.0928 a_k^2 + 0.4491 a_k - 0.0444 \qquad (54)$$

The parameter $p_k^{(c)}$ determines the deviation of the dependence of the coupling coefficient $\alpha_{010}^{(k,k-1)}$ on $a_k$ from the law $a_k^3$, $p_k^{(s)}$ - the deviation of the dependence of the resonator frequency shift due the hole in the $k$-disk on $a_k$ from the law $a_k^3$ (see (53)).

The equation (18) after separation of the real and imaginary parts and making some transformations takes the form

$$\overline{b}_k^4 - \overline{b}_k^2 - \frac{\overline{b}_k^3 \left[ \left| \rho_k^{(1)} \right| \cos(\varphi_k) - \overline{p}_{k+1} \right]}{\left| \rho_k^{(1)} \right| \sin(\varphi_k) Q_k} +$$
$$+ u_k \frac{\left[ \left| \rho_k^{(1)} \right| \left( \sin(\varphi_k + \varphi_{k-1}) - \sin(\varphi_k) \overline{p}_k \left| \rho_{k-1}^{(1)} \right| \right) - \sin(\varphi_{k-1}) \overline{p}_{k+1} \right]}{\sin(\varphi_k) \left| \rho_k^{(1)} \right| \left| \rho_{k-1}^{(1)} \right| \overline{d}_k} = 0 \qquad (55)$$

$$u_{k+1} = u_k \frac{\sin(\varphi_{k-1})}{\left| \rho_k^{(1)} \right| \left| \rho_{k-1}^{(1)} \right| \sin(\varphi_k)} - \frac{\overline{b}_k^3 \overline{d}}{\left| \rho_k^{(1)} \right| \sin(\varphi_k) Q_k}, \quad 2 \le k \le N-1 \qquad (56)$$

Parameters of the first and last resonators can be found from equations

$$\overline{b}_1^4 - \overline{b}_1^2 - \frac{u_2}{\overline{d}_1} \overline{p}_2 + \frac{u_2}{\overline{d}_1} \left| \rho_1^{(1)} \right| \cos(\varphi_1^{(1)}) = 0 \qquad (57)$$

---

[5] For simplicity, we will consider the case without of the rounding of the disk hole edges. For taking into account the rounding of the disk hole edges.

$$\beta_1 = 1 + \frac{u_2}{\bar{b}_1^{\,3} \bar{d}_1} Q_1 \left|\rho_1^{(1)}\right| \sin(\varphi_1^{(1)}) \qquad (58)$$

$$\bar{b}_N^4 - \bar{b}_N^2 - \frac{u_N}{\bar{d}_N \left|\rho_{N-1}^{(1)}\right|} \left[ \bar{p}_N \left|\rho_{N-1}^{(1)}\right| - \cos(\varphi_{N-1}^{(1)}) \right] = 0 \qquad (59)$$

$$\beta_N = \frac{u_N}{\bar{d}_N \bar{b}_N^3 \left|\rho_{N-1}^{(1)}\right|} Q_N \sin(\varphi_{N-1}^{(1)}) - 1 \qquad (60)$$

By specifying the values of the multipliers $\rho_k^{(1)}$ and a certain set of resonator parameters, from equations (55)-(60) we can find the missing set of parameters. Amplitudes $e_{010}^{(k)}$ in the chain with this full set of resonator parameters will distribute along structure in accordance with the formula (33).

Proposed approach can be used for developing of different inhomogeneous DLWs.

Let's consider the case that was studied in the work [47]. There the possibility of matching two DLWs with the same phase shift per cell $\varphi = \omega(d+t)/c = 2\pi/3$, but different aperture sizes ($a_I = $ 1.4 cm, $a_{II} = 1.3$ cm), was studied. Consideration was carried out for frequency $f = $ 2856 MHz. It was considered the case when the two different uniform DLWs ($Q = \infty$, $d = 3.0989$ cm, $t = 0.4$ cm,) were connected through one nonsymmetric transition cell. Results of calculation on the base the CCM[6] [47] for different geometries are presented in Table1 ($\Delta\varphi$ - the additional phase shift on the transition, $R$ - the reflection coefficient). Geometry obtained on the basis of equations (55)-(56) was calculated for $\rho_k^{(1)} = \exp(i\varphi)$, $k \neq s$ and $\rho_s^{(1)} = 1.146 \exp(i\varphi)$, $k = s$. The amplitude of the multiplier $\rho_s^{(1)}$ was chosen from the fact that we want to match the DLWs with $a_I = $ 1.4 cm and $a_{II} = 1.3$ cm.

Table 1

| | Geometry considered in the work [47] | | Geometry calculated on the basis of equations (55)-(56) | | Improved geometry | |
|---|---|---|---|---|---|---|
| | $a_k$ | $b_k$ | $a_k$ | $b_k$ | $a_k$ | $b_k$ |
| k=s-3 | 1.4 | 4.16874 | 1.4 | 4.16885 | 1.4 | 4.16874 |
| k=s-2 | 1.4 | 4.16874 | 1.4 | 4.16885 | 1.4 | 4.16874 |
| k=s-1 | 1.4 | 4.16874 | 1.4 | 4.16337 | 1.4 | 4.16337 |
| k=s | 1.3705 | 4.15053 | 1.34898 | 4.14577 | 1.34898 | 4.14577 |
| k=s+1 | 1.3 | 4.14066 | 1.3 | 4.14074 | 1.3 | 4.14066 |
| k=s+2 | 1.3 | 4.14066 | 1.3 | 4.14074 | 1.3 | 4.14066 |
| k=s+3 | 1.3 | 4.14066 | 1.3 | 4.14074 | 1.3 | 4.14066 |
| Results of calculation on the base the CCM | | | | | | |
| | $R = 2.5 \cdot 10^{-4}$ | $\Delta\varphi \simeq 5°$ | | | $R = 1.4 \cdot 10^{-3}$ | $\Delta\varphi \simeq 0.15°$ |

From Table 1 it follows that calculations on the base of equations (55)-(56) give the sizes of homogeneous resonators that are in good coincidence with the ones that were obtained on the basis of the rigorous electrodynamic approach (compare $b_k$ in the third and the fifth columns). The mistakes in the resonator radii are about 1 micrometer. Transition cell sizes ($a_s$, $b_s$) in the work [47] were chosen by making the reflection coefficient small enough. It was achieved by solving the diffraction problem several ten times with different $a_s$ and $b_s$. Geometry, calculated on the basis of equations (55)-(56), automatically includes in the tuning process two cells. As we see from Table 1, changing the three geometrical sizes ($a_s$, $b_s$, $b_{s-1}$) gives the better phase distribution. Finding the minimum value of the reflection coefficient by exhaustion of three sizes ($a_s$, $b_s$, $b_{s-1}$) is a difficult task. It is useful to note that the obtained value of $a_s = 1.34898$ is close to such value $\sqrt{a_{s-1}a_{s+1}} = 1.34907$.

---
[6] We used code in which each resonator couples with 6 nearby resonators

Among the slow wave waveguides, the most complex structure have the ones with phase velocities that change along the longitudinal coordinate (an injector in linacs [1,64,65,66], TWT [2,6-13]). They must ensure not only the acceleration (deacceleration) of particles, but also their grouping into small bunches. Injector sections for linacs are usually designed with a constant phase shift between cells, but with a variable length of resonators. The proposed above approach gives possibility to design the structures with the inhomogeneous phase shifts.

As example, we considered the possibility of creating smooth transition between the DLW with $\varphi_1 = 14\pi/15$ and the DLW with $\varphi_2 = 2\pi/3$ ($Q = \infty$). For $f = 2856$ MHz, $d = 3.0989$ cm, $t = 0.4$ cm the phase velocity changes from $0.71c$ to $c$.

We chose two sequences for $\rho_k^{(1)}$. The first one (the sequence N1) is

$$\rho_k^{(1)} = \begin{cases} \exp(i\varphi_1), & k < s \\ \exp(i13\pi/15), & k = s \\ \exp(i12\pi/15), & k = s+1 \\ \exp(i11\pi/15), & k = s+2 \\ \exp(i\varphi_2), & k \geq s+3 \end{cases} \quad (61)$$

The second one (the sequence N2) is

$$\rho_k^{(1)} = \begin{cases} \exp(i\varphi_1), & k < s \\ 0.949\exp(i13\pi/15), & k = s \\ 0.949\exp(i12\pi/15), & k = s+1 \\ 0.949\exp(i11\pi/15), & k = s+2 \\ \exp(i\varphi_2), & k \geq s+3 \end{cases} \quad (62)$$

Geometry calculated on the basis of equations (55)-(56) are presented in Table 2. Geometry used for calculation on the base the CCM differs in the homogeneous parts less than 2 μm.

Table 2

|  | N1 | | N2 | |
| --- | --- | --- | --- | --- |
|  | $a_k$ | $b_k$ | $a_k$ | $b_k$ |
| k=s-2 | 1.4 | 4.19633 | 1.4 | 4.19633 |
| k=s-1 | 1.4 | 4.19633 | 1.4 | 4.19633 |
| k=s | 1.4 | 4.15983 | 1.4 | 4.16117 |
| k=s+1 | 1.16863 | 4.10769 | 1.18496 | 4.11500 |
| k=s+2 | 1.06051 | 4.08584 | 1.10506 | 4.09765 |
| k=s+3 | 0.99754 | 4.07353 | 1.06792 | 4.08760 |
| k=s+4 | 0.95872 | 4.06882 | 1.04010 | 4.08261 |
| k=s+5 | 0.95872 | 4.06882 | 1.04010 | 4.08261 |
|  |  |  |  |  |
|  | R=7.29E-003 |  | R=7.64E-003 |  |

Calculation results of the longitudinal electric field distribution in the resonator centres obtained on the basis of the CCM are presented in Figure 5 and Figure 6 ($s = 11$). We see that the longitudinal electric field has nearly the same phase distribution as the chosen one for the $e_{010}^{(k)}$ amplitudes. We can also see that for the same phase distribution we can create different amplitude distributions[7] which are desirable for different types of injectors - the first distribution with the increasing amplitude [66] and .the second one with the constant amplitude [65].

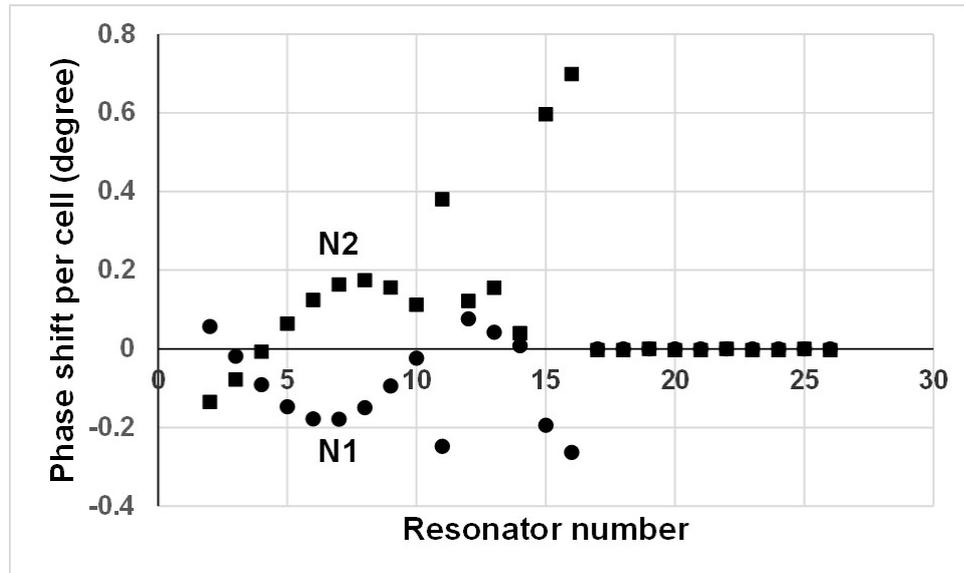

Figure 5

Difference between the specified phase shifts per cells and calculated on the basis of the CCM

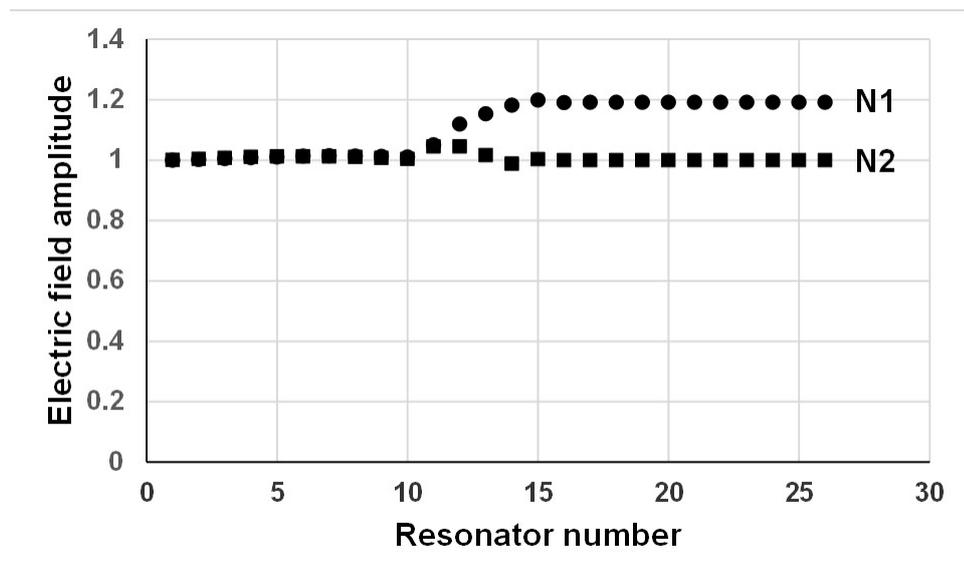

Figure 6

For high current linacs it is needed to develop accelerating sections with constant phase shifts between the cells ($\varphi_k = const$) and the amplitudes of the electric field increasing along the structure ($\left|e_{010}^{(k)}\right| \neq const$) (see, for example, [63,67]).Setting the law of amplitude variation along the structure $\left|e_{010}^{(k)}\right|$, we can find the full set of resonator parameters from equations (55)-(56) with such characteristic multipliers

---

[7] Many possible structures realize the variable phase velocity. As the power flow must be constant ($Q = \infty$), then needed distribution of electric field amplitudes determine the law of change of the aperture sizes. We can realize the increase of the phase velocity at the constant (or increasing) apertures, but the amplitudes have to increase strongly (see, for example, [8,9]).

$$\rho_k^{(1)} = \frac{\left|e_{010}^{(k+1)}\right|}{\left|e_{010}^{(k)}\right|} \exp(i\varphi) \tag{63}$$

## 4 Forward and Backward fields

In light of work on the new matrix form of second-order linear difference equations [51], we can look at the problem of expanding the electromagnetic field into the forward and backward waves in each cell of the inhomogeneous chain of resonators from the new point of view.

We have shown that in the chain that is described by the second-order difference equation (6) we can realize any reasonable ($a_n \exp(i\varphi_n)$) amplitude-phase distribution that is the product of the characteristic multipliers

$$e_{010}^{(k)} = y_k^{(1)} = y_1^{(1)} \prod_{s=1}^{k-1} \rho_s^{(1)}, \ 2 \le k \le N \tag{64}$$

For that we have to choose the resonator and opening sizes that are fulfilled the relations (55)-(56) and the parameters of couplers (57)-(60). At such geometrical sizes the second independent solution of the equation (6) equals to zero. As there is no reflection from the input coupler, we can consider that RF power transmits through the structure without reflection. This electromagnetic field we can consider as the "forward" one.

Let's suppose that the output coupler is detuned ($Z_N \rho_{N-1}^{(1)} - \alpha_{010}^{(N,N-1)} \ne 0$). What changes will occur in the distribution of the amplitudes $e_{010}^{(k)}$?

From (10),(19) it follows that the new field $\tilde{y}_k^{(2)}$ will appear in addition to the "forward" field

$$\tilde{e}_{010}^{(k)} = \tilde{y}_k^{(1)} + \tilde{y}_k^{(2)} = \tilde{y}_1^{(1)} \prod_{s=1}^{k-1} \rho_s^{(1)} + \tilde{y}_k^{(2)} \prod_{s=1}^{k-1} \rho_s^{(2)}, \ 2 \le k \le N \tag{65}$$

The amplitude of the "forward" field $\tilde{y}_1^{(1)}$ differs from the unperturbed one $y_1^{(1)}$ and depends on the initial value of the characteristic multiplier $\rho_1^{(2)}$ and the value of the output coupler detuning (see (23),(25)). The characteristic multipliers $\rho_s^{(2)}$ are the solution of the difference equation (18) with defined coefficients and with the initial value $\rho_1^{(2)}$ which we can choose arbitrary.

In the limit $Q = \infty$, when $Z_k$ is the real value and the couplers become the symmetrical elements, there is a reasonable background to consider that the amplitude of the "forward" field $\tilde{y}_1^{(1)}$ do not depends on the tuning of the output coupler. Then from (23) and (25) we obtain the initial value $\rho_1^{(2)}$

$$Z_1 - \alpha_{010}^{(1,2)} \rho_1^{(2)} = 0 \tag{66}$$

As $\beta_1 / Q_1$ has a finite value (see (39),(40)), we have

$$\left[Z_1 - \alpha_{010}^{(1,2)} \rho_1^{(1)}\right] = -2i\alpha_{010}^{(1,2)} \operatorname{Im} \rho_1^{(1)}. \tag{67}$$

From (66) it follows that

$$\rho_1^{(2)} = \operatorname{Re} \rho_1^{(1)} - i \operatorname{Im} \rho_1^{(1)} = \rho_1^{(1)*} \tag{68}$$

and, as $Z_k$ is the real value, from (18) we obtain

$$\rho_k^{(2)} = \rho_k^{(1)*}. \tag{69}$$

16Therefore, the additional field that arises due to reflection from the output coupler becomes the conventional backward field.

The problem becomes more difficult at $Q \neq \infty$. The characteristic multipliers $\rho_k^{(2)}$ that define the structure of additional field are the solution of the difference equation (18). This equation we can rewrite as

$$\rho_{k+1}^{(2)} = -\frac{\alpha_{010}^{(k+1,k)}}{\alpha_{010}^{(k+1,k+2)} \rho_k^{(2)}} + \frac{\alpha_{010}^{(k+1,k)}}{\alpha_{010}^{(k+1,k+2)} \rho_k^{(1)}} + \rho_{k+1}^{(1)}, 1 \leq k \leq N-2, \quad (70)$$

where $\rho_1^{(2)} \neq \rho_1^{(1)}$ is a free parameter. If $\alpha_{010}^{(k+1,k)} = \alpha_{010}^{(k+1,k+2)}$, $1 \leq k \leq N-2$ (the homogeneous chain) and $\rho_k^{(1)} = \exp(i\varphi - \gamma)$ ($1 \leq k \leq N-1$) we have the solution of the equation (70) in the analytical form

$$\rho_k^{(2)} = \frac{1}{\rho_k^{(1)}} = \exp(-i\varphi + \gamma), \ 1 \leq k \leq N-1. \quad (71)$$

In the general case, the equation (70) has no simple solution. As $\rho_1^{(2)}$ is a free parameter and there is not reasonable background for its choice, then the structure of additional field and its amplitude $\tilde{y}_1^{(2)}$ are not define uniquely. Moreover, the amplitude of the "forward" field $\tilde{y}_1^{(1)}$ which depend on $\rho_1^{(2)}$ and $\rho_{N-1}^{(2)}$ (see, (23)} is not define uniquely, too.

Therefore, in the frame of considered model the separation of the electromagnetic field into "forward" and "backward" components in the inhomogeneous chain of resonators is not define uniquely. It is needed to apply some additional criteria for defining the properties of "reflected" fields.

## Conclusions

We presented the novel approach to the synthesis of the electromagnetic field distribution in a chain of coupled resonators that can be described by the second-order difference equation for amplitudes of expansion of the electromagnetic field. This approach is based on the new matrix form of the solutions of the second-order difference equations that give possibility to construct the two linearly independent solutions. Setting the structure of one solution, from the Riccaty equation we can find the electrodynamical characteristics of resonators and coupling holes, at which the desired distribution of amplitudes is realized. Several examples show that proposed approach can be useful in solving different physical problems. On the base of this approach we also considered the problem of separation of the electromagnetic field into "forward" and "backward" components in the inhomogeneous chain of resonators. It was shown that in the frame of considered model such separation is not defined uniquely.

The problem of creating a special field distribution is attracting attention of different researchers. This problem arises at the construction and design of new materials including nano-materials with so called cloaking properties (see, for example, [68,69,70,71]). The proposed approach can be used as a numerical tool to design 1-D devices and materials that manipulate waves in a specified manner.

## References

1 T.P.Wangler. Principles of RF linear accelerators, John Wiley & Sons, Inc., 1998

2 A. S.Gilmour. Klystrons, Traveling Wave Tubes, Magnetrons, Crossed-Field Amplifiers, and Gyrotrons, ARTECH HOUSE, 2011

3 I.C. Hunter. Theory and design of microwave filters, The Institution of Engineering and Technology, 2006